\NewCommandCopy{\LaTeXtextbf}{\textbf}
\documentclass{ieeeaccess}
\RenewCommandCopy{\textbf}{\LaTeXtextbf}

\usepackage{xcolor}
\usepackage{amsmath}
\usepackage{url}
\usepackage{graphicx}
\usepackage{multirow}
\usepackage{xspace}
\usepackage{lipsum}
\usepackage{listings}
\usepackage[linesnumbered,ruled]{algorithm2e}

\SetCommentSty{mycommfont}

\colorlet{punct}{red!60!black}
\definecolor{background}{HTML}{EEEEEE}
\definecolor{delim}{RGB}{20,105,176}
\colorlet{numb}{magenta!60!black}

\lstdefinelanguage{json}{
    basicstyle=\scriptsize\ttfamily,
    showstringspaces=false,
    breaklines=true,
    tabsize=2,
    backgroundcolor=\color{background},
    literate=
     *{:}{{{\color{punct}{:}}}}{1}
      {,}{{{\color{punct}{,}}}}{1}
      {\{}{{{\color{delim}{\{}}}}{1}
      {\}}{{{\color{delim}{\}}}}}{1}
      {[}{{{\color{delim}{[}}}}{1}
      {]}{{{\color{delim}{]}}}}{1}
      {.}{{{\color{punct}{.}}}}{1}
      {\ \ }{\ }1,
}

\newcommand{\acro}{\ensuremath{\sf{QED}}\xspace}
\newcommand{\pone}{\ensuremath{\sf{P1}}\xspace}
\newcommand{\ptwo}{\ensuremath{\sf{P2}}\xspace}
\newcommand{\pthree}{\ensuremath{\sf{P3}}\xspace}

\newcommand{\ignore}[1]{}

\newlength\mylen
\newcommand\myinput[1]{%
  \settowidth\mylen{\KwIn{}}%
  \setlength\hangindent{\mylen}%
  \hspace*{\mylen}#1\\}

\SetKw{Continue}{continue}
\SetKw{Is}{is}
\SetKw{Not}{not}

\begin{document}
\history{This work has been submitted to the IEEE for possible publication. Copyright may be transferred without notice, after which this version may no longer be accessible.}
\doi{10.1109/ACCESS.2024.DOI}

\title{A Toolchain for Assisting Migration of Software Executables Towards Post-Quantum Cryptography}

\author{Norrathep Rattanavipanon\authorrefmark{1}, 
Jakapan Suaboot\authorrefmark{1}, and Warodom Werapun\authorrefmark{1}}
\address[1]{College of Computing, Prince of Songkla University, Phuket Campus (e-mail: norrathep.r@psu.ac.th, jakapan.su@psu.ac.th, warodom.w@psu.ac.th)}
\tfootnote{This work was supported by
the National Science, Research and Innovation Fund (NSRF)
and Prince of Songkla University (Grant No. COC6701016S).}

\corresp{Corresponding author: Jakapan Suaboot (e-mail: jakapan.su@psu.ac.th).}

\begin{abstract}
Quantum computing poses a significant global threat to today's security mechanisms. 
As a result, security experts and public sectors have issued guidelines to help organizations migrate their software to post-quantum cryptography (PQC). Despite these efforts, there is a lack of (semi-)automatic tools to support this transition especially when software is used and deployed as binary executables.
To address this gap, in this work, we first propose a set of requirements necessary for a tool to detect quantum-vulnerable software executables.
Following these requirements, we introduce \acro: a toolchain for \underline{Q}uantum-vulnerable \underline{E}xecutable \underline{D}etection.
\acro uses a three-phase approach to identify quantum-vulnerable dependencies in a given set of executables, from file-level to API-level, and finally, precise identification of a static trace that triggers a quantum-vulnerable API.
We evaluate \acro on both a synthetic dataset with four cryptography libraries and a real-world dataset with over 200 software executables.
The results demonstrate that: (1) \acro discerns quantum-vulnerable from quantum-safe executables with 100\% accuracy in the synthetic dataset; (2) \acro is practical and scalable, completing analyses on average in less than 4 seconds per real-world executable; and (3) \acro reduces the manual workload required by analysts to identify quantum-vulnerable executables in the real-world dataset by more than 90\%.
We hope that \acro can become a crucial tool to facilitate the transition to PQC, particularly for small and medium-sized businesses with limited resources.

\end{abstract}


\titlepgskip=-21pt

\maketitle

\section{Introduction}
Advancements in quantum computing are accelerating as demonstrated by multiple recent breakthroughs.
From two qubits in 1998~\cite{chuang1998experimental}, quantum computers are now capable of supporting over 1,000 qubits~\cite{ibm2023}. Although the current number of qubits (and the error-correcting capability) are still insufficient for practical use, 
the current pace of quantum computing development along with substantial investment from major tech companies~\cite{ibm2023,microsoft,google} leads many experts to believe that a large fully-functional quantum computer will become commercially viable within the next two decades~\cite{nist}.

The distinct advantage of quantum computers lies in their ability to leverage quantum mechanics to solve certain mathematical problems that are not feasible with classical computers. 
This capability has several societal implications. Quantum computers could potentially benefit certain tasks, e.g., optimizing traffic systems~\cite{neukart2017traffic}, enhancing machine learning~\cite{schuld2015introduction}, and accelerating the discovery of new drugs and materials~\cite{cao2018potential}. 
However, there are also potential negative impacts.
Malcious actors could exploit quantum computers to undermine today's widely-used security mechanisms, particularly public-key cryptosystems\footnote{This in turn may lead to attacks on symmetric-key cryptosystems through ``harvest now, decrypt later" strategy~\cite{harvestnowdecryplater}.}.
For instance, breaking RSA cryptosystem could be done with 4,098 logical (noise-free) qubits of a quantum computer~\cite{haner2016factoring}, while elliptic curves would need 2,330 logical qubits~\cite{roetteler2017quantum}.
%

With increasing awareness of future quantum attacks, global concerns about security have intensified. In response, researchers and public sectors are urging all organizations to prepare for these threats~\cite{joseph2022transitioning,nist-quantum-ready,ott2019identifying,pandey2023cryptographic}. 
For instance, NIST~\cite{nist-quantum-ready} recommends establishing quantum-readiness teams within an organization to prepare migration of their software systems to post-quantum cryptography (PQC). 
This preparation includes: (1) creating cryptographic inventories to assess the use of cryptography within the organization, and (2) conducting risk assessments based on these inventories.

Despite existing recommendations and guidelines, there are currently no (semi-)automatic tools specifically designed for assisting in this task.
Consequently, in practice, analysts must rely on a mix of various scattered tools and manual analysis to identify software systems that are vulnerable to quantum attacks.
This results in significant and tedious effort, especially given the potential large amount of software systems within organizations. 
Moreover, the analysts often lack access to the source code and therefore must perform PQC migration based solely on program binaries. This in turn requires advanced knowledge in binary analysis, which translates to increased costs in terms of budget, time, and labor, making it unaffordable for small and medium-sized businesses.

Motivated by these challenges, we first formulate the requirements for a tool to assist in migration of software executables towards PQC.
After establishing these, we introduce \underline{Q}uantum-vulnerable \underline{E}xecutable \underline{D}etection (\acro), a toolchain designed for detecting quantum-vulnerable executables. 
The main goal of \acro is to aid analysts in identifying quantum-vulnerable (QV) executables located in a specific computer/server witin an organization.  
The design rationale of \acro revolves around two key observations. 

First, as a security practice, software applications typically do not implement cryptographic functionalities within themselves but rather utilize them via API access to well-established cryptography libraries (e.g., OpenSSL~\cite{openssl}, wolfSSL~\cite{wolfssl}, MbedTLS~\cite{mbedtls}).
Hence, QV executables can be determined based on existence of the dependencies with QV APIs provided by these libraries.
Second, since most programs either are non-cryptographic or use relatively quantum-safe cryptography (e.g., cryptographic hash functions), \acro should employ a fast analysis to eliminate this type of executables in the earlier phases.
This allows for an accurate but more resource-intensive analysis to be performed on a smaller set of more-likely-to-be-QV candidates later. To achieve this, we design \acro to operate in three phases where:

\begin{itemize}
    \item \acro's first phase aims to determine software executables that have dependency on QV cryptography libraries.
    \item The second phase identifies QV APIs within those libraries and eliminates executables that have no dependency on these APIs.
    \item The final phase pinpoints QV executables by performing a static callgraph analysis to identify a trace of function calls from the executable's entry (i.e., \verb|main| function) to a QV API.
\end{itemize}

Each phase of \acro provides a human-readable report, informing analysts about post-quantum risk assessment of each program executable. This information can then be used for decision-making in later PQC migration plans within organizations.

\textbf{Contribution:} In this work, we aim to make the following contributions:

\begin{enumerate}
    \item We outline the requirements of a tool aimed at assisting in a PQC migration task of software executables, with a specific focus on identifying QV software executables.

    \item We introduce a toolchain, called \acro, designed to meet the proposed requirements. 
    \acro implementation is open-sourced and publicly available at~\cite{repo}.

    \item We empirically validate accuracy and efficiency of \acro using both synthetic and real-world datasets. The results show that \acro produces only one false negative while achieving a 100\% true positive rate and reducing the analyst's manual workload in identifying QV software by over 90\%.
\end{enumerate}

Overall, we hope \acro can become a crucial tool in easing the transition of existing software systems to PQC, especially for small and medium-sized organizations that lack the labor and resources needed for PQC migration.


\section{Related Work}\label{sec:related}

\subsection{Cryptographic Discovery Tools in Software Executables}

Existing work leverages either static or dynamic analysis to discover cryptography usage in software binaries.
Static analysis based approaches typically rely on known static features, such as initialization vectors, look-up tables (e.g., S-Boxes) or a sequence of assembly instructions, to detect cryptography implementation in executables~\cite{benedetti2017detection,findcrypt2,kanal,lestringant2015automated}. 
These approaches are lightweight in nature but may miss detecting some cryptography functions when executables are optimized by an aggresive compiler, which causes a modification to the target static features in binaries.

Another line of approaches utilizes dynamic analysis to overcome this limitation. 
One common method is to identify cryptography usage based on runtime information, e.g., the presence of loops~\cite{lutz2008towards}, loop structures~\cite{xu2017cryptographic}, or the input-output relationship of a function~\cite{grobert2011automated,zhao2011detection}, collected from execution traces.
In practice, generating meaningful runtime execution traces can be challenging, especially when the analyst has no access to the source code or is not fully familiar with the software. 
To address this challenge, recent approaches have proposed using dynamic features created through symbolic execution, eliminating the need for actual runtime execution traces.
For example, ALICE~\cite{eldefrawy2020towards} and Harm-DOS~\cite{weideman2022harm} perform symbolic execution on binary functions to determine whether their output matches the expected hash output, thereby implementing the expected hash function, without running the entire software.
Additionally, the work in~\cite{meijer2021s} extends the concept of Data Flow Graph isomorphism~\cite{lestringant2015automated} with symbolic analysis to detect proprietary cryptography implementations in software binaries.

Our approach falls into the category of static analysis. However, unlike existing static approaches, our method detects cryptography usage based on the API name information, which is always present in binaries despite aggressive compiler optimizations, hence incurring no false negatives.
Compared to existing dynamic approaches, \acro does not require running executables to collect execution traces or performing heavyweight symbolic analysis. Moreover, none of the existing tools focuses on identifying \emph{quantum-vulnerable} cryptography usage in executables or consider a scenario where cryptography implementations are dynamically linked with executables.

\subsection{Transitioning Organizations to PQC}
Migrating organizations towards PQC is currently an ongoing global effort.
Related work in this area~\cite{nist-quantum-ready, lamacchia2021long,nather2024migrating, IEEE-access-pqc-migrate} has focused on providing advisories and guidelines for PQC migration to organizations. At high-level, the migration process can be structured into four steps~\cite{nather2024migrating}:

\begin{enumerate}
    \item \textbf{Diagnosis.} Identify QV cryptography used within an organization.
    \item \textbf{Planning.} Use the information collected from the previous step to create a plan and a timeline for migration.
    \item \textbf{Execution.} Perform the migration according to the previously identified plan.
    \item \textbf{Maintenance.} Monitor changes within the organization after migration execution to maintain PQC.
\end{enumerate}

Our work falls into the first step that aims to reduce manual workloads required by organizations to identify software executables vulnerable to quantum attacks. The results produced by our tool can be subsequently used in the later steps to execute and maintain the PQC migration plans.

Another closely related concept is crypto agility~\cite{alnahawi2023state}, which refers to the property of a system that can be easily adapted to a different cryptographic algorithm. In contrast to our work, this concept involves the second and third steps of the migration process. We expect that after identifying QV (and perhaps crypto-unagile) software systems using our tool, the organization can replace them with quantum-safe and crypto-agile alternatives.

The migration process can also be categorized based on where QV cryptography is used: (1) in compiled binary executables or their dependencies, (2) in assets on external/end-user systems, and (3) in the network communication layer~\cite{nist-quantum-ready,hasan2024framework}. Our work targets the first type of usage, while the others are potential avenues for future research.

Major cryptography libraries are in the process of incorporating PQC into their implementations, such as Open Quantum Safe OpenSSL~\cite{openssl-qs} and Bouncy Castle~\cite{bouncy}. However, as these implementations are still under active development, they are not yet ready for deployment in production software~\cite{nather2024migrating}.
To the best of our knowledge, there is currently no other concrete, publicly available tool specifically geared towards the identification of software binaries that utilize QV APIs from cryptography libraries. We believe our tool will be useful in reducing the cost of PQC transition, especially for small to medium-sized organizations.

\section{Problem Statement}\label{sec:scope}

In this work, we envision a scenario where an analyst wants to take the first step towards PQC migration by identifying QV software within an organization.
The organization may consist of computers/servers, each containing several software systems.
Within the scope of this work, we consider the computers to be Linux machines with software programs written in \verb|C| or \verb|C++|, deployed as Linux executables in the Executable and Linkable Format (ELF). 
Nonetheless, the overall idea in this work can be extended to programs written in other languages and running under different operating systems.

We assume that a software executable utilizes cryptography by accessing APIs provided by common cryptography libraries through dynamic linking.
This means, software executables that implement cryptographic functions within themselves or statically link with cryptography libraries are considered out of scope.
This assumption aligns with the standard security and program linking practices for Linux machines~\cite{collberg2005slinky}.
 
We define a software executable as QV if there is at least one possible execution path from its entry point (i.e., the start address of the \verb|main| function) to one of the cryptography library's APIs implementing QV algorithms (e.g., RSA, Diffie-Hellman, Elliptic Curve-based Digital Signatures).
This API may reside within the same executable or in a shared library used at runtime by this executable.
Then, given a list of software executables, we consider a tool effective if it can identify all executables that are QV according to this definition, along with human-understandable evidence supporting their classification.


\section{Tool Requirements}\label{sec:req}

We now formulate the requirements for a tool designed to address the above-mentioned scenario.

\begin{itemize}
    \item[\textbf{\fbox{R1}}] \textbf{Dynamic Linking.} The tool must be able to identify executables that use QV APIs via dynamic linking.
    
    \item[\textbf{\fbox{R2}}] \textbf{Binary-level Analysis.} The tool must not rely on the availability of executables' source code in any part of its analysis. This is because programs deployed in an organization may be developed by third parties who are unwilling to share/publish the source code due to intellectual property concerns.
    
    \item[\textbf{\fbox{R3}}] \textbf{Static-only Features.} The tool must operate using only static features of executables without making use of dynamic features (e.g., runtime execution traces). In practice, an organization could contain too many software executables for analysts to know how to meaningfully run each of them in order to generate the required dynamic features.
    
    \item[\textbf{\fbox{R4}}] \textbf{Scalability.} The tool must support the analysis for a large number of software executables while being able to complete its analysis within a reasonable runtime, on the order of minutes. 
    
    \item[\textbf{\fbox{R5}}] \textbf{Effectiveness.} 
    As the main goal is to reduce the analyst's workload by minimizing the number of executables required for manual analysis, the tool must incur no false negatives, i.e., it should effectively identify all QV executables. While high accuracy is crucial, the tool can tolerate a small false positive rate, i.e., it may classify a small set of executables as QV when they are not. The task of analyzing and eliminating these false positives is manageable when they are few and can be handled through the analyst's manual review.
\end{itemize}

\begin{figure*}
    \centering
    \includegraphics[width=\textwidth]{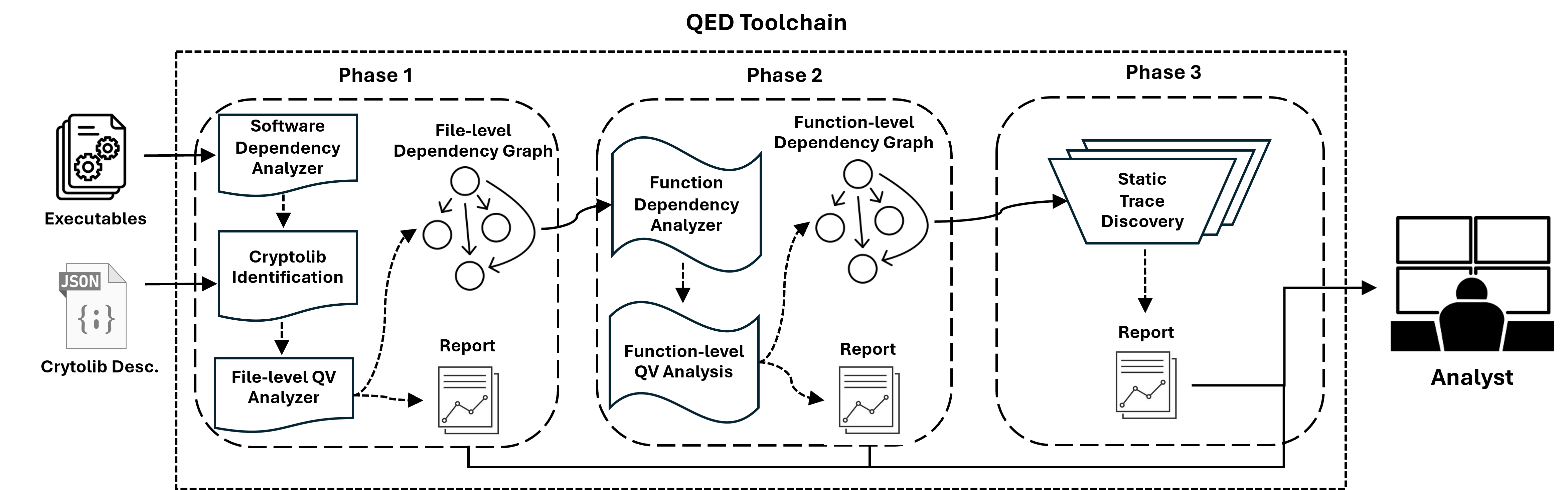}
    \caption{\acro Components}
    \label{fig:design}
\end{figure*}

\section{\acro Toolchain}\label{sec:design}


Figure~\ref{fig:design} overviews the workflow in the \acro toolchain. 
It takes two inputs: (i) a list of software executables and (ii) descriptions of QV cryptography libraries.
For (i), the analyst can supply a directory path storing executables (e.g., \verb|/bin/|) to \acro;
whereas, the second input (ii) can be created using a list of QV APIs that uniquely identifies the library of interest. 
For example, to describe OpenSSL 1.1~\cite{openssl}, the analyst creates a list of its QV APIs, i.e., \verb|{RSA_Sign, EC_Sign, ..., DH_compute_key}| as a JSON file (see Cryptolib Desc. in Figure~\ref{fig:design}) and feeds it to \acro.
This list can be compiled manually by reviewing all available APIs and identifying the ones that are QV, or semi-automatically by first filtering APIs based on specific QV keywords (e.g., \verb|RSA|, \verb|EC| and \verb|DH|) and then manually analyzing them.

\acro consists of three phases where each utilizes increasingly complex analysis that in turn yields more accurate results. 
All phases rely on only static analysis, eliminating the need to understand the context of executables to run them.
At the end of each phase, \acro produces human-readable reports that describe the risk level associated with quantum threats for each executable, along with supporting evidence. 
The analyst can then choose which report to use based on whether speed or accuracy is prioritized. If accuracy is favored, the analyst can work with the output from the final phase, which takes the longest to run but yields fewest false positives. Meanwhile, if speed is more important, the analyst can stop \acro at an earlier phase and use its reports instead.  

\subsection{Phase 1: File-level Dependency Analysis}

\begin{algorithm}[htp]
  \DontPrintSemicolon
  \SetKwFunction{algo}{P1}
  \SetKwFunction{genSwGraph}{GenSWDepGraph}
  \SetKwFunction{genSwGraphHelper}{GenSWDepGraphHelper}
  \SetKwFunction{findCryptoLibs}{FindCryptoLibs}

  \SetKwProg{myalg}{Algorithm}{}{}
  \SetKwProg{myproc}{Procedure}{}{}
  
  \KwIn{$execs$ -- list of software executables}
  \myinput{$descs$ -- list of crypto library's descriptions}
  \KwOut{$G_{1}, EV_1$}
  
  \myalg{\algo{$execs, descs$}}{
      $G_{1} \leftarrow$ \genSwGraph{execs}\;
      $cryptoLibs \leftarrow$ \findCryptoLibs{$G_{1}.nodes$}\;
      \ForEach{$n \in G_{1}.nodes$}{
        \If{$\neg~G_{1}.hasPath(n, cryptoLibs)$}{
            $G_{1}.removeNode(n)$\;
        }
      }
      $EV_{1} \leftarrow \{\}$ \;
      \ForEach{$e \in execs$}{
        $EV_{1}.append(G_{1}.paths(e,cryptoLibs))$\;
      }
      \KwRet $G_{1}, EV_{1}, cryptoLibs$\;
  }~\\

  \myproc{\genSwGraph{$execs$}}{
    \tcp{Initialize $G$ as empty directed graph}
    $G \leftarrow $ \texttt{DiGraph()} \;
    \ForEach{$e \in execs$}{
        $G \leftarrow$ \genSwGraphHelper{$G, e$}\;
    }
    \KwRet $G$\;
  }~\\

  \myproc{\genSwGraphHelper{$G, e$}}{
    \If{$e \in G.nodes$}{
        \KwRet $G$
    }
    $G.addNode(e)$ \;
    $libs \leftarrow $ \texttt{GetDirectDep}($e$) \;
    \If{$libs$ \Is $\phi$}{
        \KwRet $G$
    }
    \ForEach{$l \in libs$}{
        $G.addEdge(e, l)$ \;
        $G \leftarrow$ \genSwGraphHelper{$G,l$} \;
    }
    \KwRet $G$\;
  }~\\

  \myproc{\findCryptoLibs{$nodes, descs$}}{
    $cryptoLibs \leftarrow \{\}$ \;
    \ForEach{$n \in nodes$}{
        $f \leftarrow$ \texttt{GetExportFuncs}($n$) \;
        \If{$\exists~d \in descs$ s.t. $d \subseteq f$} {
            $cryptoLibs.append(d)$ \;
        }
    }
    \nl \KwRet $cryptoLibs$\;
  }~\\

  \caption{File-level Dependency Analysis (\pone)}
  \label{alg:p1}
\end{algorithm} 

Given a list of executables ($execs$) and QV APIs exported by crytography libraries ($descs$), the first phase (\pone) of \acro aims to perform a static analysis at file-level to identify executables that have (direct or indirect) dependency on these libraries.
Details of this phase are shown in Algorithm~\ref{alg:p1}.

To achieve this goal, we first model software dependency of $execs$ as a directed graph $G_1 = (V_1, E_1)$, where $V_1$ is the set of files (including $execs$ and shared objects that they depend on) and $E_1$ is the set of tuples describing a file-level direct dependency relationship.
For example, $(n_1, n_2) \in E_1$ means that file $n_1$ directly depends on file $n_2$, i.e., $n_1$ contains a call to one of the functions exported by $n_2$.

\pone generates $G_1$ by performing a depth-first search to discover all dependencies of $execs$, as described in \texttt{GenSWDepGraph} of Algorithm~\ref{alg:p1}.
Once $G_1$ is constructed, \pone locates QV cryptography libraries in $V_1$.
This is done in \texttt{FindCryptoLibs} of Algorithm~\ref{alg:p1} by analyzing exported functions of each node and checking whether these functions match with one of crypto libraries' description in $descs$.
When they do, it considers the node to implement the same QV APIs, and therefore most likely to be the target cryptography library.

After discovering all cryptography libraries ($cryptoLibs$) in $V_1$, \pone prunes out the nodes that have no dependency on $cryptoLibs$, which is done by checking whether a path from a node to any of $cryptoLibs$ exists (Lines 5-9 in Algorithm~\ref{alg:p1}).
As a result, \pone passes a refined $G_1$, consisting only of executables with file-level dependencies on QV APIs, to \acro's next phase.

\textbf{Evidence.} \pone reports to analysts $EV_1$ consisting of file-level dependency paths from each executable in $execs$ to a cryptography library.
An example of $EV_1$ is shown in Figure~\ref{fig:evidence}.

\textbf{Accuracy.} \pone yields no false negatives: suppose executable $e$ is QV according to the definition in Section~\ref{sec:scope}. This means either $e$ or one of its shared libraries must invoke one of the QV APIs. 
The former case implies $e$ has a direct dependency on the cryptography library while the latter results in an indirect dependency through the shared library. 
Therefore, in either case, there must be a file-level dependency between $e$ and a cryptography library, making $e$ appear in $G_1$ and $EV_1$.


Nonetheless, \pone incurs substantial false positives since it may incorrectly report executables that depend on a cryptography library but utilize none of its QV APIs (e.g., it uses the quantum-safe \verb|SHA512| algorithm from OpenSSL).


\begin{figure*}

\begin{minipage}{.33\textwidth}
    \begin{lstlisting}[language=json]
"EV_1": [
    {
        "path": [
            "/usr/bin/sftp",
            "/usr/lib/libcrypto.so.1.1"
        ]
    }
    ,
    {
        "path": [
            "/usr/bin/dig",
            /usr/lib/libdns.so",
            "/usr/lib/libcrypto.so.1.1"
        ]
    }
    ,
    {
        "path": [
            "/usr/bin/nmap",
            "/usr/lib/libssl.so.1.1",
            "/usr/lib/libcrypto.so.1.1"
        ]
    }
    ...
    {
        "path": [
            "/usr/bin/curl",
            "/usr/lib/libcurl.so.4",
            "/usr/lib/libcrypto.so.1.1"
        ]
    }
]
\end{lstlisting}

\end{minipage}\hfill
\begin{minipage}{.33\textwidth}
    \begin{lstlisting}[language=json]
"EV_2": [
    {
        "path": [
            "/usr/bin/nmap",
            "/usr/lib/libssl.so.1.1",
            "/usr/lib/libcrypto.so.1.1"
        ],
        "QV_apis": [
            "DSA_do_sign",
            "DSA_do_verify",
            "EVP_PKEY_get1_DSA",
            ...
            "RSA_verify"
        ]
    }
    ...
    {
        "path": [
            "/usr/bin/curl",
            "/usr/lib/libcurl.so.4",
            "/usr/lib/libcrypto.so.1.1"
        ],
        "QV_apis": [
            "DH_get0_key",
            "DSA_get0_key",
            "DSA_get0_pqg",
            "EVP_PKEY_get0_DH",
            ...
            "RSA_get0_key"
        ]
    }
]
\end{lstlisting}
\end{minipage}\hfill
\begin{minipage}{.33\textwidth}
    \begin{lstlisting}[language=json]
"EV_3": [
    {
        "static-trace": [
            [
                "/usr/bin/nmap",
                "main"
            ],
            [
                "/usr/bin/nmap",
                "sub_3f340"
            ]
            ...
            [
                "/usr/bin/nmap",
                "SSL_CTX_new"
            ],
            [
                "/usr/lib/libssl.so.1.1",
                "SSL_CTX_new"
            ]
            ...
            [
                "/usr/lib/libssl.so.1.1",
                "EVP_PKEY_get0_RSA"
            ],
            [
                "/usr/lib/libcrypto.so.1.1",
                "EVP_PKEY_get0_RSA"
            ]
        ]
    }
]
\end{lstlisting}
\end{minipage}
    \caption{Evidence reports produced by \acro's first phase (left), second phase (middle) and third phase (right)}
    \label{fig:evidence}
\end{figure*}

\subsection{Phase 2: API-level Dependency Analysis}

\begin{algorithm}[htp!]
  \DontPrintSemicolon
  \SetKwFunction{algo}{P2}

  \SetKwProg{myalg}{Algorithm}{}{}
  \SetKwProg{myproc}{Procedure}{}{}
  
  \KwIn{$G_1, execs, cryptoLibs$}
  \KwOut{$G_2, EV_2$}
  
  \myalg{\algo{$G_1, execs, cryptoLibs$}}{
        $G_2 \leftarrow G_1$\;
        \ForEach{$c \in cryptoLibs$}{
            $pred \leftarrow G_2.predecessors(c)$\;
            \ForEach{$p \in pred$}{
                $f \leftarrow \texttt{GetExternFuncs}(p)$\;
                \If{$f \cap c.desc = \phi$} {
                    $G_2.removeEdge(p, c)$\;
                }
            }
        }
        
        \ForEach{$n \in G_2.nodes$}{
            \If{\Not $G_2.hasPath(n, cryptoLibs)$}{
                $G_2.removeNode(n)$\;
            }
        }
        
        \ForEach{$(src,dst) \in G_2.edges$}{
            $apis \leftarrow \texttt{GetExternFuncs}(src) \cap \texttt{GetExportFuncs}(dst)$ \;
            $G_2.setAPI((src,dst),apis)$\;
        }
        
        $EV_2 \leftarrow \{\}$ \;
        \ForEach{$e \in execs$}{
            $EV_2.append(G_2.paths(e,cryptoLibs))$\;
        }
        \KwRet $G_2, EV_2$\;
  }

  \caption{API-level Dependency Analysis (\ptwo)}
  \label{alg:p2}
\end{algorithm} 

To reduce the false positives in \pone, \acro's second phase (\ptwo) analyzes executables' dependencies at API-level.
To achieve this, \ptwo constructs an API dependency graph $G_2 = (V_2, E_2)$, where $V_2$ is the set of files, similar to $V_1$, while $E_2$ corresponds to the set of three-element tuples describing an API-level dependency relationship, e.g., $(n_1, n_2, apis) \in E_2$ indicates file $n_1$ invokes external functions specified in a list of APIs, $apis$, which are implemented and exported by file $n_2$. This analysis is detailed in Algorithm~\ref{alg:p2}.

\ptwo starts by initializing $G_2$ as a copy of $G_1$ and determining predecessor nodes of $cryptoLibs$, i.e., those with direct dependencies on $cryptoLibs$.
Then, it checks whether these nodes contain function calls to QV APIs, specified in $descs$. If no such calls exist, it identifies the node as a false positive and its API-level dependency is excluded in $G_2$ by removing the edge connecting this node and the library node.

Next, \ptwo locates other nodes that do not depend on predecessors of $cryptoLibs$ by checking for the absence of a path between these nodes and any $cryptoLibs$ nodes in $G_2$ (Lines 12-16 in Algorithm~\ref{alg:p2}).
Once these nodes are identified, \ptwo removes them, leaving $G_2$ with only nodes that have dependencies on QV predecessors.

After that, \ptwo goes through each directed edge in $G_2$ and aims to embed it with a set of APIs used/exported by the connecting nodes.
For an edge from node $n_1$ to node $n_2$,
\ptwo identifies a set of external APIs used in $n_1$ and a set of APIs exported by $n_2$ (found in \verb|.dynsym| section of $n_1$ and $n_2$ ELF executables, respectively).
It then intersects these sets to find the APIs invoked by $n_1$ that belong to $n_2$, and updates the edge with this API-level dependency information (Lines 17-20 in Algorithm~\ref{alg:p2}). It then passes $G_2$ to the next next phase.

\textbf{Evidence.} 
\ptwo outputs $EV_2$ to analysts. Similar to $EV_1$, $EV_2$ contains file-level dependency paths where the first element is in $execs$ and the last is in $cryptoLibs$. Unlike $EV_1$, $EV_2$ ensures that a predecessor of $cryptoLibs$ (the second last element) in this path contains a call to QV APIs, which are also included in $EV_2$ (see Figure~\ref{fig:evidence}).
It also implies that when an executable $e$ has a direct dependency on one of $cryptoLibs$ (i.e., $e$ is itself a predecessor of the library), $EV_2$ guarantees $e$ to contain invocations to QV APIs.

\textbf{Accuracy.} \ptwo also avoids false negatives for the similar reason as in \pone. However, it addresses false positives in which $e$ exclusively uses quantum-safe APIs from $cryptoLibs$ without relying on other QV shared libraries. Nevertheless, false positives may still occur in two scenarios:

\begin{enumerate}
    \item $e$ contains invocations to external functions provided by a shared library (different from $cryptoLibs$), where these functions do not use QV APIs but other functions exported by the same shared library do. In this case, \ptwo still classifies $e$ as QV since it propagates QV detection from the shared library, even though $e$ does not use any of its QV exported functions.

    \item $e$ may contain calls to QV APIs, but this function call could be inside ``dead'' code that is never executed at runtime.
    In this scenario, \ptwo incorrectly considers $e$ as QV since it can only determine whether these calls exist in $e$, but not whether they will actually be executed at runtime.
\end{enumerate}

\subsection{Phase 3: Static Trace Analysis}

\begin{algorithm}[!htp]
  \DontPrintSemicolon
  \SetKwFunction{algo}{P3}
  \SetKwFunction{getStaticTrace}{GetStaticTrace}

  \SetKwProg{myalg}{Algorithm}{}{}
  \SetKwProg{myproc}{Procedure}{}{}
  
  \KwIn{$G_2, EV_2, conservative$}
  \KwOut{$EV_3$}
  
  \myalg{\algo{$G_2, EV_2$}}{
        $EV_3 \leftarrow \{\}$\;
        \ForEach{$path \in EV_2$}{
            $mainAddr \leftarrow \texttt{GetMainAddress}(path[0])$\;
            $trace \leftarrow \getStaticTrace(G_2, path, 0, mainAddr)$\;
            \uIf{$trace$ \Is \Not $\phi$}{
                $EV_3.append(trace)$\;
            }\ElseIf{conservative}{
                $EV_3.append(path)$\;
            }
        }
        \KwRet $EV_3$\;
  }~\\
  
  \myproc{\getStaticTrace{$G, path, i, fromFunc$}}{
    $apis \leftarrow G.getAPI((path[i], path[i{+}1]))$\;
    $CG \leftarrow \texttt{GenCallgraph}(path[i])$\;
    \ForEach{$toFunc \in apis$}{
        $t1 \leftarrow CG.path(fromFunc, toFunc)$\;
        \If{$i$ \Is $len(path){-}1$}{
            \KwRet $t1$\;
        }
        \If{$t1$ \Is \Not $\phi$}{
            $t2 \leftarrow \getStaticTrace(G, path, i{+}1, toFunc)$\;
            \If{$t2$ \Is \Not $\phi$}{
                \KwRet $t1 || t2$\;
            }
        }
    }
    \KwRet $\phi$\;
  }

  \caption{Static Trace Analysis (\pthree)}
  \label{alg:p3}
\end{algorithm}

The third phase (\pthree) of \acro takes $G_2$ and $EV_2$ as input and aims to precisely identify executables that conform to the QV definition in Section~\ref{sec:scope}.
To achieve this, \pthree must return evidence $EV_3$ such that each of its entries corresponds to an execution trace starting from the executable entry point (e.g., \verb|main| address) to a QV API call. 
This gives assurance to the analyst that the executable is QV, as each entry in $EV_3$ serves as proof of a QV API's execution for a given executable.

\pthree supports two modes: normal and conservative. In normal mode, the absence of an execution trace directly indicates the executable is non-QV. 
In contrast, conservative mode treats missing traces as potential false negatives and returns the evidence from the previous phase (i.e., $EV_2$) to the analyst for further manual verification.
These modes offer a trade-off between minimizing false negatives and reducing manual effort, allowing the analysts to select based on their specific use-cases.

Recall that a path in $EV_2$ corresponds to a list of files: $[n_1, n_2, ..., n_k]$, where $n_1 \in execs$, $n_k \in cryptoLibs$ and $n_i$ has direct dependency on $n_{i+1}$. In the special case of $n_{k-1}$, it additionally indicates that $n_{k-1}$ contains a call to a QV API provided by $cryptoLibs$.
To create $EV_3$, \pthree first analyzes each file for each path in $EV_2$ using \texttt{GetStaticTrace} of Algorithm~\ref{alg:p3}.
For the first file, $n_1$, it enumerates $n_2$'s external functions (stored in $E_2$) and, for each $n_2$'s external function $f_{ex}$, checks whether a call to $f_{ex}$ is reachable from the executable entry point, i.e., the \verb|main| function in this case.
\pthree performs this check by constructing a static callgraph of $n_1$, represented as a directed graph with functions as nodes and invocations as edges.
Then, it validates reachability by checking whether a path from the \verb|main| node to the $f_{ex}$ node exists within this callgraph.

If reachability is not found, \pthree concludes that $f_{ex}$ cannot be accessed by $n_1$'s runtime execution and thus moves on the analysis to the next $n_2$'s external function.
Upon discovering reachability, \pthree repeats the analysis on the next file $n_2$, i.e., by constructing $n_2$'s callgraph and performing the reachability analysis with $f_{ex}$ as the entry point and $n_3$'s external function as the end point.
This process continues until it finds a reachable path in $n_{k-1}$, as indicated in Algorithm~\ref{alg:p3}.

\textbf{Evidence.} \pthree generates a static trace in $EV_3$ by concatenating all paths returned by the reachability analyses from files $n_1$, ..., $n_{k-1}$, an example of which is shown in Figure~\ref{fig:evidence}. When \pthree fails to find a static trace for a specific executable and is configured as the conservative mode, it considers this executable a false negative and returns the corresponding entry from $EV_2$ for that executable (Lines 8-10).

\textbf{Accuracy.} $EV_3$ minimizes false positives since an entry in $EV_3$ represents a static execution trace from an executable entry point to a QV API call, demonstrating how this QV API can be invoked at runtime. 
Nonetheless, \pthree's reachability analysis may incur false negatives 
as constructing a complete callgraph for \verb|C| programs is still considered an open problem~\cite{shoshitaishvili2016sok}.
To avoid false negatives, $EV_3$ may fall back to the results from $EV_2$ in such cases (as discussed above), prompting the analyst to perform manual analysis on them later.

\section{Evaluation}\label{sec:eval}

\subsection{Implementation \& Experimental Setup}\label{sec:impl}

We implemented the \acro toolchain using $\approx$800 lines of Python3 code. 
It leverages pyelftools\footnote{https://github.com/eliben/pyelftools} library to detect ELF files and extract relevant information, such as dynamic symbols as well as exported and external functions, from the detected ELF files.
Directed graphs in \acro are built and manipulated  using NetworkX library\footnote{https://networkx.org/}.
Meanwhile, \acro's \pthree constructs static callgraphs from executables via the angr framework~\cite{shoshitaishvili2016sok}. 
We conducted experiments with \acro on an Ubuntu 20.04 system running atop an Intel i5-8520U @ 1.6GHz with 24GB of RAM.
The source code for \acro and its evaluated datasets are publicly available at~\cite{repo}.

\begin{figure} [ht]
    \centering
    \includegraphics[width=0.8\columnwidth]{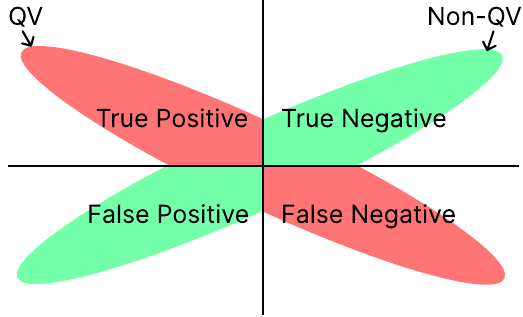}
    \caption{QV and non-QV prediction outcomes}
    \label{fig:tpfp-classification}
\end{figure}

To measure \acro's accuracy, we use true positive rate (TPR) and true negative rate (TNR) metrics. A true positive (TP) occurs when \acro correctly identifies a QV executable, while a false positive (FP) happens when \acro incorrectly classifies a non-QV executable as QV. True negatives (TN) and false negatives (FN) are defined similarly. For clarity, these prediction outcomes are illustrated in Figure~\ref{fig:tpfp-classification}.

The main challenge in evaluating \acro's accuracy is the difficulty of precisely identifying false negatives (i.e., QV executables that \acro might miss) incurred by \acro. In real-world datasets, each binary executable may depend on multiple complex shared libraries. Thus, to accurately identify all false negatives in this case, it would require manually checking each shared library for QV API usage (in addition to the application binaries);
this represents a large amount of manual effort that becomes practically infeasible as the number of software executables in the dataset grows.

To address this, we evaluated \acro using two datasets: synthetic and real-world. The synthetic dataset was specifically crafted to allow us to determine the ground truth (i.e., whether an executable is QV or non-QV) of executables, enabling us to precisely measure \acro's accuracy (including false negatives).
In contrast, our evaluation of the real-world dataset focuses more on \acro's scalability and quantifies the reduction in manual workload saved by \acro.

\subsection{Synthetic Dataset: Description}\label{sec:synthetic_desc}

The synthetic dataset includes four cryptography libraries: OpenSSL v1.1.1\footnote{\url{https://github.com/openssl/openssl/tree/OpenSSL_1_1_1f}}, OpenSSL v3.3.1\footnote{\url{https://github.com/openssl/openssl/tree/openssl-3.3.1}; note that we distinguish between OpenSSL v3.3 and v1.1, as many APIs from OpenSSL v1.1 are deprecated in v3.3.}, MbedTLS v2.28.8\footnote{\url{https://github.com/Mbed-TLS/mbedtls/tree/v2.28.8}}, and wolfSSL v5.7.2\footnote{\url{https://github.com/wolfSSL/wolfssl/tree/v5.7.2-stable}}.
To create this dataset, we ported example programs provided by the respective libraries, where each program directly calls the library APIs to implement a specific cryptographic primitive. We selected five primitives (and thus five programs in total) for each library: SHA-512, AES-256, Diffie-Hellman Key Exchange, RSA, and ECDSA, with only the last three being QV in this dataset.
This amounts $5 \times 4 = 20$ executables in the first set. As the programs in this set utilize the library APIs directly within their executable, we call this set ``Direct Dependency''.

Further, to test \acro against indirect usage of QV APIs, we developed a second set of C programs that have indirect dependency on the crypto library, i.e., they do not directly call the crypto library APIs but rather access them through a separate shared library; as such, we denote this set ``Indirect Dependency''.
To implement this set, we created a single shared library exporting five APIs, each implementing one of the aforementioned cryptographic primitives. We then constructed five C programs, each dynamically linking with this shared library to invoke one of its exported APIs. 
As with the first set, we utilized the same four cryptography libraries, resulting in $5 \times 4 = 20$ executables in the second set.
In total, the synthetic dataset comprises 40 executables with 24 QV and 16 non-QV executables, as summarized in Table~\ref{tab:synthetic}.

\begin{table}[]
\centering
\caption{Breakdown of number of executables in synthetic dataset}
\label{tab:synthetic}
\resizebox{.8\columnwidth}{!}{%
\begin{tabular}{l||c c c}
\hline
Synthetic Dataset & QV & Non-QV & Total \\ \hline\hline
Direct Dependency      & 12  & 8      & 20    \\ \hline
Indirect Dependency    & 12  & 8      & 20    \\ \hline\hline
Total             & 24 & 16     & 40    \\ \hline
\end{tabular}%
}
\end{table}

\subsection{Synthetic Dataset: Accuracy Results}\label{sec:synthetic_res}



\begin{table*}[!htp]
\centering
\caption{Accuracy of \acro evaluated on synthetic dataset, where \# QV (\# Non-QV) represents the number of QV (non-QV) executables input to each phase and TPR/TNR refer to the true positive/negative rate.}
\label{tab:synthetic-accuracy}
\resizebox{.9\textwidth}{!}{%
\begin{tabular}{l||cc|cc|cc}
\hline
\multicolumn{1}{c||}{\multirow{2}{*}{\begin{tabular}[c]{@{}l@{}}\acro's Phases ($\rightarrow$)\\ Synthetic Dataset ($\downarrow$)\end{tabular}}}   & \multicolumn{2}{c|}{\pone}                & \multicolumn{2}{c|}{\pone+\ptwo}                & \multicolumn{2}{c}{\pone+\ptwo+\pthree}                \\
                    & TP/FN (TPR) & TN/FP (TNR) & TP/FN (TPR) & TN/FP (TNR) & TP/FN (TPR) & TN/FP (TNR) \\ \hline\hline
Direct Dependency  & 12/0 (100\%)  &  0/8 (0\%)  & 12/0 (100\%) & 8/0 (100\%) &  12/0 (100\%) & 8/0 (100\%)    \\ \hline
Indirect Dependency  & 12/0 (100\%)   &  0/8 (0\%)  & 12/0 (100\%) & 0/8 (0\%) & 12/0 (100\%) & 8/0 (100\%) \\ \hline\hline
Total               & 24/0 (100\%)   &  0/16 (0\%)     & 24/0 (100\%)  & 8/8 (50\%) & 24/0 (100\%) & 16/0 (100\%)  \\ \hline
\end{tabular}%
}
\end{table*}

Table~\ref{tab:synthetic-accuracy} depicts the accuracy of \acro at each phase.
The results show that \pone identifies all executables in the synthetic dataset as positives (or QV) because all of them have file-level dependencies on cryptography libraries. This leads to a 100\% TPR, but also in the misclassification of 16 non-QV executables, resulting in a 0\% TNR.

After running \ptwo, \acro removes all false negatives (i.e., 8 non-QV executables) in the Direct Dependency set in which, despite dynamically linking with cryptography libraries, they exclusively use their non-QV APIs. 
However, \ptwo still fails to identify non-QV executables in the Indirect Dependency set because it cannot track API usages across executables. Despite this, TNR improves to 50\% in \ptwo while maintaining a perfect TPR.
Finally, by performing static trace analysis, \pthree can discover the remaining false positives and eliminates all of them, acheiving 100\% TPR and TNR.

\begin{table*}[]
\centering
\caption{Statistics of real-world dataset; ``\# Execs'' represents the number of executables included in each set, ``\# Unique Deps'' is the number of unique dependencies required for each set and ``Avg. \# Deps'' is the average number of dependencies per executable}
\label{tab:rw-stats}
\resizebox{.6\linewidth}{!}{%
\begin{tabular}{l||c c c c c}
\hline
\multirow{2}{*}{Set}    & \multirow{2}{*}{\# Execs} & Avg. Exec & \# Unique      & Avg.      & Avg. Dep  \\ 
                        &          & Size (in KB)   & Deps   & \# Deps         &  Size (in KB) \\ \hline \hline
Coreutils   & 109 &  355 & 7 & 1.52   & 492  \\ \hline
UnixBench           & 18 & 19 & 2 & 1.06  & 1660 \\ \hline
Network        & 13 & 950 & 76 & 13.38 & 898                       \\ \hline 
tpm2-tools          &  86 & 53 & 48 & 9.45    & 433   \\ \hline \hline
Total              & 226 & 248 & 84 & 5.19 & 834                       \\ \hline
\end{tabular}%
}
\end{table*}

\subsection{Real-world Dataset: Description}\label{sec:rw-desc}

In the second dataset, we evaluated \acro on real-world software executables. Specifically, we selected four sets of Linux software in this dataset: 
\begin{itemize}
    \item \textbf{Coreutils}~\cite{coreutils} and \textbf{UnixBench}~\cite{unixbench} are non-cryptographic software commonly used to benchmark the performance of Unix-like machines.
    Since all executables in this set are non-cryptographic, they are considered to be non-QV.
    \item \textbf{Network} includes 13 popular Linux networking utility programs, including curl, dig, netcat, nmap, nslookup, ping, scp, sftp, ssh, sshd, telnet, tracepath and wget.
    Due to the small number, we managed to manually inspect their source codes/binaries to determine the ground truth. Out of these, we found 7 programs to be QV as they rely on QV APIs provided by OpenSSL v1.1.
    \item \textbf{tpm2-tools}~\cite{tpm2} implements TPM functionalities in software and uses cryptographic features through the tpm2-tss library\footnote{https://github.com/tpm2-software/tpm2-tss}, which in turn calls the cryptography APIs from OpenSSL v1.1.
    Unlike the previous set, this set contains a large number of programs (86), making it impractical to determine their ground truths manually.
\end{itemize}
Table~\ref{tab:rw-stats} summarizes the statistics of all sets in our real-world dataset. In total, this dataset contains 226 executables with an average size of 248KB and each depending on an average of 5 shared libraries with an average size of 834KB. The descriptions of this dataset are reported in our open-source repository~\cite{repo}.

\subsection{Real-world Dataset: Results}\label{sec:rw-results}


\begin{table*}[]
\centering
\caption{Accuracy of \acro evaluated on real-world dataset, where n/a means the analysis in a specific phase was not conducted since \acro completed the analysis in the earlier phase.
tpm2-tools is excluded since we cannot determine their ground truth reliably.}
\label{tab:realworld-accuracy}
\resizebox{.9\textwidth}{!}{%
\begin{tabular}{l||cc|cc|cc}
\hline
\multicolumn{1}{c||}{\multirow{2}{*}{\begin{tabular}[c]{@{}l@{}} Phases ($\rightarrow$)\\ Set ($\downarrow$)\end{tabular}}}   & \multicolumn{2}{c|}{\pone}                & \multicolumn{2}{c|}{\pone+\ptwo}                & \multicolumn{2}{c}{\pone+\ptwo+\pthree}                \\
                     & TP/FN (TPR) & TN/FP (TNR) & TP/FN (TPR) & TN/FP (TNR) & TP/FN (TPR) & TN/FP (TNR) \\ \hline\hline
Coreutils   & 0/0 (100\%)     & 109/0 (100\%)       & n/a      & n/a    & n/a   & n/a     \\ \hline
UnixBench   & 0/0 (100\%)     & 18/0 (100\%)       & n/a      & n/a    & n/a   & n/a     \\ \hline
Network     & 7/0 (100\%)     &  4/2 (67\%)    & 7/0 (100\%)     & 6/0 (100\%)  & 6/1 (86\%)  & 6/0 (100\%)    \\ \hline
\end{tabular}%
}
\end{table*}

\subsubsection{Accuracy}
As mentioned in Section~\ref{sec:impl}, our evaluation of the real-world dataset focuses on performance/manual effort reduction rather than accuracy, as obtaining the ground truth of large-scale real-world executables is practically infeasible. However, for the sake of completeness, we report the accuracy results on the Coreutils, UnixBench, and Network sets, where their ground truths can be reliably determined. The results for these sets are summarized in Table~\ref{tab:realworld-accuracy}. 

As non-cryptographic software, Coreutils and UnixBench have no dependencies on any cryptography library, they are non-QV.
\acro identifies these and ends its analyses at \pone without the need to further run \ptwo and \pthree, yielding 100\% TPR and TNR.
In contrast, the Network set potentially relies on cryptography implementations from OpenSSL. In this set, \pone classifies 9 programs as QV, passing them to the subsequent phases. Of these, 2 are non-QV, resulting in 2 false positives for \pone;
these false positives are then detected and excluded in \ptwo.
We manually inspect these false positives and found that they are caused by sfp and scp programs that dynamically link with OpenSSL but invoke only its non-QV APIs (e.g., \verb|RAND_bytes|);
this is not detectable in \pone.

More importantly, the results in Table~\ref{tab:realworld-accuracy} support \acro's design choice that aims to minimize false negatives. Both \pone and \ptwo produce no false negatives, and only one false negative of the curl program is reported by \pthree in the Network set. 
Upon closer inspection, this false negative occurs because, although curl uses QV APIs from OpenSSL, its usage is based on indirect calls (e.g., through function pointers), which are notoriously difficult to detect through static callgraph analysis that is currently considered an open challenge~\cite{shoshitaishvili2016sok}.
Nonetheless, we emphasize that if avoiding any false negatives is critical, the analyst can opt for the conservative version of \pthree. This version would classify curl as QV, providing its $EV_2$ evidence to the analyst for further manual analysis to verify its QV status.







\subsubsection{Performance}

\begin{figure*}[!htp]
\centering
\caption{Runtime to complete each phase of \acro on real-world dataset, where ``ALL'' contains all sets, i.e., Coreutils+UnixBench+Network+tpm2-tools}
\includegraphics[width=.8\linewidth]{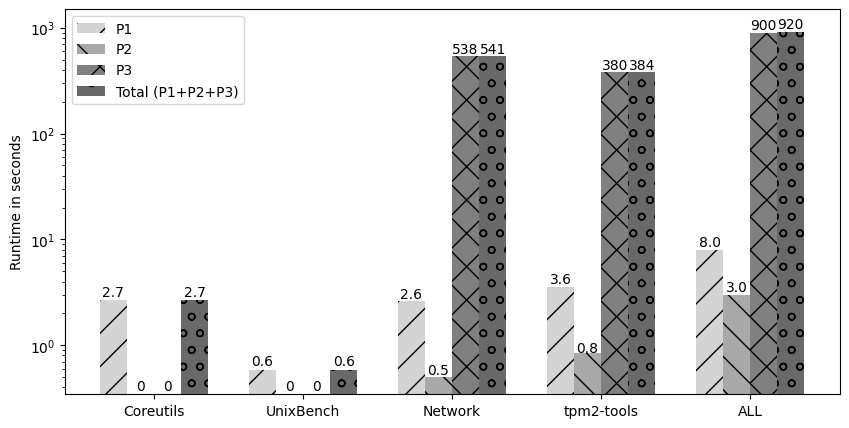}
\label{fig:rw-runtime}
\end{figure*}

Runtime results of \acro on the real-world dataset are reported in Figure~\ref{fig:rw-runtime}.
As expected by \acro's design, \pone and \ptwo complete their analyses very quickly (under 8 seconds) across all sets.
In contrast, as \pthree involves expensive static callgraph analysis, it takes significantly more time and thus dominates \acro's overall runtime.
Overall, \acro takes around 15 minutes to process all 226 executables, resulting in an average of 4 seconds per executable.

\begin{table}[]
\centering
\caption{Callgraph construction (C.C.) time for the tpm2-tools and Network set}
\resizebox{\linewidth}{!}{%
\begin{tabular}{l||cc|cc}
\hline
\multirow{2}{*}{Set}                       & \multicolumn{2}{c|}{tpm2\--tools} & \multicolumn{2}{c}{Network} \\
                                           & Execs         & Deps         & Execs           & Deps          \\ \hline\hline
\# files analyzed                      & 86            & 3            & 7               & 8             \\ \hline
Avg. size (in KB)                     & 54            & 543          & 1574            & 596           \\ \hline
Avg. C.C. Time (in sec)  & 2.7           & 38           & 34              & 37            \\ \hline
Total C.C. Time (in sec) & 232           & 114          & 238             & 296           \\ \hline
\end{tabular}
}
\label{tab:callgraph-time}
\end{table}

\begin{figure}[!htp]
\centering
\caption{Time to construct callgraph of each executable and dependency in Network and tpm2-tools}
\includegraphics[width=\linewidth]{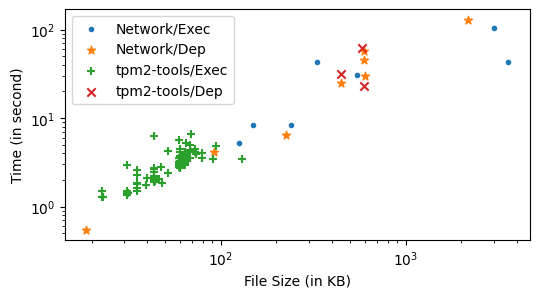}
\label{fig:cc-runtime}
\end{figure}

To better understand how the size of analyzed files affects \pthree's runtime, we measured the time required to construct a static callgraph (representing the main overhead in \pthree) for each analyzed file in the Network and tpm2-tools sets. 
The results shown in Table~\ref{tab:callgraph-time} indicate that while \pthree processes 70 more files in Network than in tpm2-tools, the majority of the analyzed files in tpm2-tools are much smaller, averaging around 54KB for 86 executables. Consequently, \pthree can build callgraphs of these smaller files very quickly, under 3 seconds.

However, the Network set encompasses 15 files that are over 500KB in size, leading to more than 10 times longer for the callgraph construction time. This relationship is further illustrated in Figure~\ref{fig:cc-runtime}, which shows the time required to construct a callgraph for each individual file. The figure confirms that larger files require more time to process, as they tend to contain more functions, resulting in larger and more complex callgraphs.

\begin{table}[!htp]
\centering
\caption{Peak RAM usage (in MB) of \acro on real-world dataset}
\resizebox{.8\linewidth}{!}{%
\begin{tabular}{l || c c c}
\hline
Phase        & \pone & \pone+\ptwo  & \pone+\ptwo+\pthree   \\ \hline\hline
Coreutils  & 179 & n/a & n/a  \\ \hline
UnixBench  & 179 & n/a & n/a  \\ \hline
Network    & 181 & 182 & 4,958 \\ \hline
tpm2-tools & 183 & 187 & 3,561 \\ \hline\hline
All        & 184 & 190 & 6,791 \\ \hline
\end{tabular}
}
\label{tab:rw-ram}
\end{table}

We report the peak RAM usage in Table~\ref{tab:rw-ram}. Consistent with the runtime results, \pone and \ptwo use a small amount of RAM, around 180MB. In contrast, \pthree requires significantly more memory, ranging from 3 to 5GB. Despite this higher memory requirement, \pthree's RAM usage remains within a reasonable range that can easily be acquired in standard, commodity computers.

\subsubsection{Manual workload reduction}
\label{sec:eval-manual}

\begin{table*}[!htp]
\centering
\caption{Manual effort reduction of \acro on real-world dataset; note that \pthree is a conservative one; the real-world dataset contains a total of 226 executables.}
\resizebox{.65\linewidth}{!}{%
\begin{tabular}{l||ccc}
\hline
Phase  & \pone      & \pone + \ptwo      & \pone + \ptwo + \pthree     \\ \hline\hline
\# execs required further manual review             & 95      & 93      & 20      \\ \hline
\% of manual workload reduction & 57.96\% & 58.84\% & 91.15\% \\ \hline
\end{tabular}
}
\label{tab:rw-saving}
\end{table*}

Lastly, we quantify the reduction in manual workload as the number of files requiring further manual analysis to identify their QV status. 
We note that the standard \pthree produces $EV_3$ evidence, which includes a static QV execution trace. While this necessitates some manual checking to confirm its presence in the binaries, it is negligible compared to the effort needed to manually analyze $EV_1$ or $EV_2$, which involve reverse engineering binaries or source code (if available). 
Therefore, we exclude the standard \pthree from this evaluation. 

Instead, we consider \acro with the conservative version of \pthree, which guarantees no false negatives by falling back to $EV_2$ evidence for executables where no static QV execution trace is found. 
Unlike the standard \pthree, the analyst must manually review this $EV_2$ for its QV API usage.

Table~\ref{tab:rw-saving} illustrates the manual workload reduction achieved by \acro. With just \pone and \ptwo, \acro can already reduce the analyst's manual workload by over 50\%. Incorporating the conservative version of \pthree further reduces the manual effort, leaving only 20 possible QV executables ($<10\%$) for manual inspection.



\section{Meeting the Requirements}

We argue that \acro satisfies all requirements in Section~\ref{sec:req} via the descriptions of \acro's design in Section~\ref{sec:design} and experimental results in Section~\ref{sec:eval} as follows:

\begin{itemize}
    \item[\textbf{\fbox{R1}}] \textbf{Dynamic Linking.} \acro performs file-dependency analysis in \pone to determine all dependencies of the executables, including shared cryptography libraries. 
    This information is subsequently used in \ptwo and \pthree to accurately detect whether these executables not only dynamically link with the cryptography library but also directly or indirectly invoke its QV APIs.
    
    \item[\textbf{\fbox{R2}}] \textbf{Binary-level Analysis.} All phases in \acro operate on the binary executable level and do not require the corresponding source code at any point during its analyses.
    
    \item[\textbf{\fbox{R3}}] \textbf{Static-only Features.} \pone and \ptwo require extracting dynamic symbols from the executables, while \pthree necessitates constructing static call graphs. All of these features are static and can be found in the executables without the need to run them.
    
    \item[\textbf{\fbox{R4}}] \textbf{Scalability.} Our experimental results demonstrate that \acro completes its analyses on 226 executables under 10 minutes, averaging about 4 seconds per executable, which is practically reasonable.
    
    \item[\textbf{\fbox{R5}}] \textbf{Effectiveness.} 
    \acro is designed to minimize the number of false negatives, as supported by our experimental results: out of all tested executables, only one is classified as a false negative by \pthree, while \pone and \ptwo incur no false negatives, albeit with higher false positive rates. 
    We emphasize that if avoiding false negatives is an important concern, the conservative version of \pthree can be employed, though this would increase the analyst's manual workload to identify and eliminate the resulting false positives.
\end{itemize}

\section{Conclusion and Future Work}\label{sec:conclusion}
We identified the requirements for a tool to assist in migrating software executables towards post-quantum cryptography. Based on these requirements, we developed a toolchain called \acro that detects quantum-vulnerable executables. It operates in three phases, each employing increasingly sophisticated techniques.
Its design rationale is to quickly locate and eliminate quantum-safe software using fast analyses in the earlier phases, while applying more precise but resource-intensive methods in the final phase. We evaluated \acro on both synthetic and real-world datasets, demonstrating its effectiveness, scalability, and ability to reduce manual workload required for analysts to migrate their organizations towards post-quantum cryptography.

\acro has some limitations that suggest avenues for future research. First, as a static analysis tool, \acro cannot detect quantum-vulnerable API usages arising from indirect calls (i.e., function pointers), leading to potential false negatives. A possible future direction is to incorporate lightweight dynamic analysis to identify these indirect calls and hence eliminate such false negatives.
Second, \acro currently assumes executables are dynamically linked with quantum-vulnerable cryptography libraries. As a result, it does not account for scenarios where executables implement cryptographic functions within themselves or statically link with cryptography libraries. Extending \acro to detect cryptographic functions directly within executables could address this limitation.
Lastly, \acro could be expanded beyond merely identifying quantum-vulnerable usages to also (semi-)automatically patch them with quantum-safe and crypto-agile implementations.



\appendices
\section{\break Data Availability}
The source code of the proposed \acro is accessible through the GitHub repository at: https://github.com/norrathep/qed.git.

\bibliographystyle{ieeetr}
\bibliography{main}
\vspace{-10 mm}
\begin{IEEEbiography}
[{\includegraphics[width=1in,height=1.25in,clip,keepaspectratio]{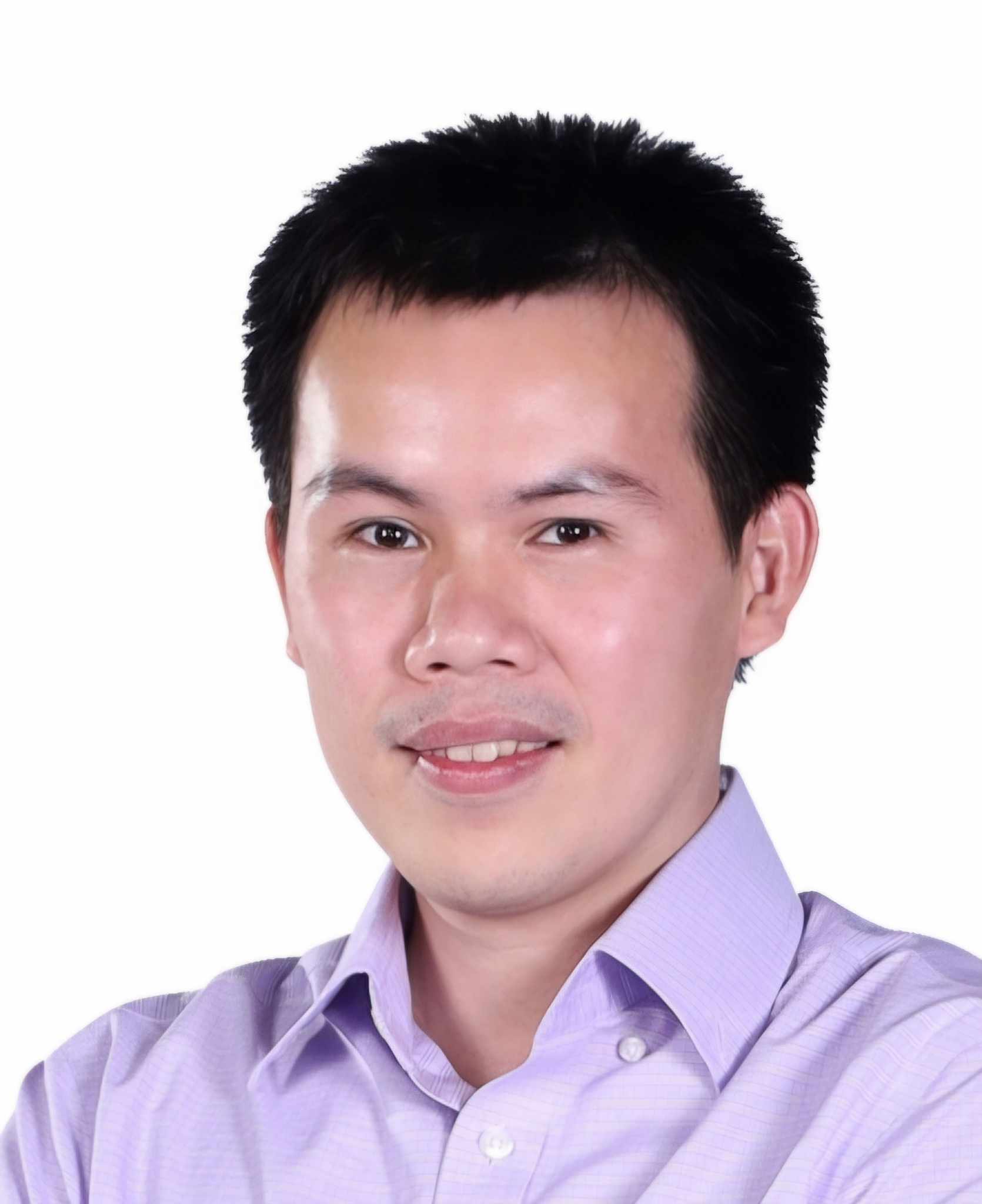}}]{Norrathep Rattanavipanon} 
received his Ph.D. in Computer Science from the University of California, Irvine in 2019. Currently, he is an assistant professor with the College of Computing, Prince of Songkla University, Phuket Campus. 
His research interests lie in the area of security and privacy, particularly in post-quantum cryptography, embedded systems and IoT security, software and binary analysis, and security/privacy in machine learning systems.
He has served as a technical program committee at top-tier security conferences: NDSS and ACM CCS.
His work has appeared at leading security and hardware venues such as USENIX Security, ACM CCS, IEEE/ACM DAC and IEEE/ACM ICCAD.
\end{IEEEbiography}
\vspace{-10 mm}
\begin{IEEEbiography}[{\includegraphics[width=1in,height=1.25in,clip,keepaspectratio]{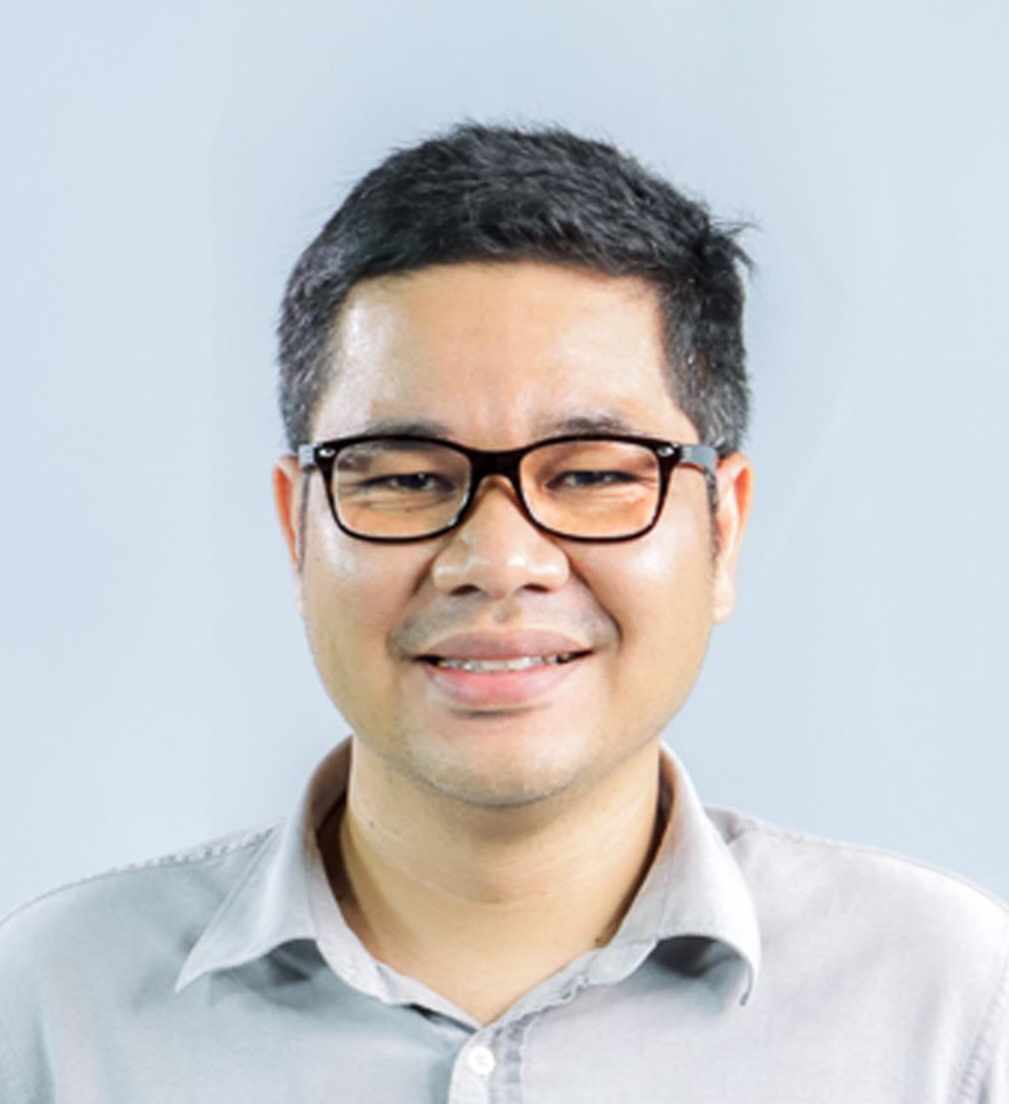}}]{Jakapan Suaboot}received his B.Eng and M.Eng (research) degrees in Computer Engineering from Prince of Songkla University (Thailand), and his Ph.D. from RMIT University (Australia) in 2007, 2010, and 2021, respectively. He is a lecturer at the College of Computing, Prince of Songkla University, Phuket Campus. His research interests include malware detection, data breach prevention, machine learning technologies, and DeFi security.
\end{IEEEbiography}
\vspace{-10 mm}
\begin{IEEEbiography}[{\includegraphics[width=1in,height=1.25in,clip,keepaspectratio]{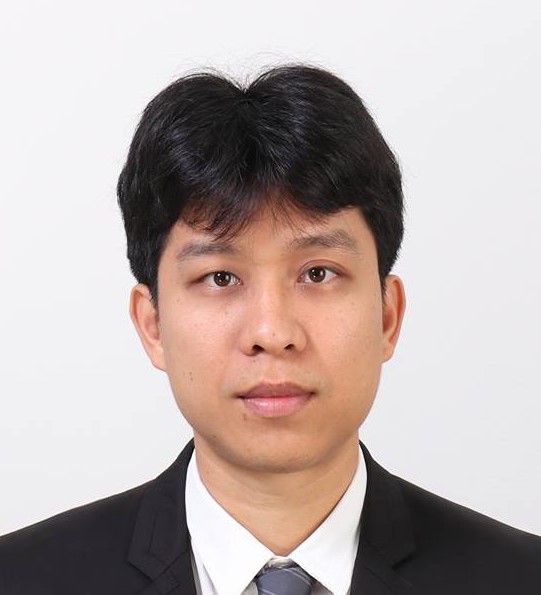}}]{Warodom Werapun} received the Ph.D. degree in Computer Engineering from ENSEEIHT/INP University, France in 2012. He is an assistant professor, the Head of the BLOCK Research Team, and the Associate Dean for Research and Innovation of the College of Computing, Prince of Songkla University, Phuket Campus. He has received 4 awards in blockchain-related projects at national and international levels. His research interests include blockchain, smart contracts, web programming and cybersecurity. 
\end{IEEEbiography}

\EOD

\end{document}